\documentclass[aip,jcp,amsmath,amssymb, reprint,]{revtex4-1}

\usepackage{graphicx}
\usepackage{dcolumn}
\usepackage{siunitx}

\usepackage{bm}

\usepackage[a-1b]{pdfx} 
\usepackage[utf8]{inputenc}
\usepackage[T1]{fontenc}
\usepackage{mathptmx}
\usepackage{listings}
\usepackage{rotating} 

\usepackage{color}
\usepackage{dcolumn} 
\usepackage{bm} 
\usepackage{graphicx}
\usepackage{multirow} 
\usepackage{pifont} 
\usepackage{epsfig}
\usepackage{amsmath} 
\usepackage{subfigure}
\usepackage{float}
\usepackage{booktabs}
\usepackage{tabularx}
\usepackage{natbib}
\usepackage{gensymb}
\usepackage{rotating}
\usepackage[normalem]{ulem}

\begin{document}

\author{Run R. Li}
\affiliation{
             Department of Chemistry and Biochemistry,
             Florida State University,
             Tallahassee, FL 32306-4390}

\author{Nicholas C. Rubin}
\affiliation{
             Google Research, 
             Mountain View, CA, USA}
             
\author{A. Eugene DePrince III}
\affiliation{
             Department of Chemistry and Biochemistry,
             Florida State University,
             Tallahassee, FL 32306-4390}
\email{adeprince@fsu.edu}

\title{Challenges for variational reduced-density-matrix theory: Total angular momentum constraints}

\begin{abstract}

The variational two-electron reduced density matrix (v2RDM) method is generalized for the description of total angular momentum ($J$) and projection of total angular momentum ($M_{J}$) states in atomic systems described by non-relativistic Hamiltonians, and it is shown that the approach exhibits serious deficiencies. Under ensemble $N$-representability constraints, v2RDM theory fails to retain the appropriate degeneracies among various $J$ states for fixed spin ($S$) and orbital angular momentum ($L$), and, for fixed $L$, $S$, and $J$, the manifold of $M_{J}$ states are not necessarily degenerate. Moreover, a substantial energy error is observed for a system for which the two-electron reduced density matrix is exactly ensemble $N$-representable; in this case, the error stems from violations in pure-state $N$-representability conditions. Unfortunately, such violations do not appear to be good indicators of the reliability of energies from v2RDM theory in general. Several states are identified for which energy errors are near zero and yet pure-state conditions are clearly violated. 

\end{abstract}

\maketitle

\section{Introduction}

A large body of work has sought the variational determination of the elements of the two-electron reduced density matrix (2RDM),\cite{Husimi:1940:264, Lowdin:1955:1474,Mayer:1955:1579,Percus64_1756,Rosina75_868,Rosina75_221,Garrod75_300,Rosina79_1366,Erdahl79_147,Fujisawa01_8282,Mazziotti02_062511,Mazziotti06_032501,Percus04_2095,Zhao07_553,Lewin06_064101,Bultinck09_032508,DePrince16_423,DeBaerdemacker11_1235,VanNeck15_4064,DeBaerdemacker18_024105,Mazziotti17_084101,Ayers09_5558,Bultinck10_114113,Cooper11_054115,Mazziotti08_134108,DePrince16_2260,Mazziotti16_153001} without knowledge of the $N$-electron wave function, as a means of circumventing the exponential complexity of exact wave function methods. The principal difficulty in this field is that a large number of non-trivial constraints\cite{Erdahl01_042113,Mazziotti12_263002,Mazziotti16_032516,Klyachko06_72,Klyachko08_287} must be placed on the 2RDM in order guarantee that it derives from an $N$-electron density matrix (a pure state) or an ensemble of $N$-electron density matrices (an ensemble state). Nevertheless, algorithmic advances\cite{Mazziotti04_213001,Mazziotti11_083001} have made large-scale applications\cite{Mazziotti11_5632,Mazziotti17_3142,Mazziotti17_9377,Mazziotti18_4988,DePrince19_6164} of variational 2RDM (v2RDM) theory that enforce approximate ensemble $N$-representability conditions somewhat commonplace. Enforcing pure-state conditions, while more difficult, is also possible,\cite{DePrince16_164109} with similar algorithms. These practical advances notwithstanding, v2RDM theory is beset by a variety of unsolved theoretical challenges. Two such problems include (i) the tendency of the approach to dissociate heteronuclear diatomic molecules into fractionally-charged species\cite{Ayers09_5558,Bultinck10_114113,Cooper11_054115} (stemming from a derivative discontinuity issue similar to that which arises in density functional theory {\color{black}[DFT]}\cite{Ayers00_5172,Yang08_792}) and (ii) an apparent inability to model angular momentum projection states in a balanced way\cite{Ayers12_014110,DePrince19_032509} {\color{black}(which, again, is an problem reminiscent of one that plagues DFT\cite{Savin96_327})}. This contribution focuses on the latter issue.

It has long been recognized that the application of suitable spin constraints\cite{Mazziotti05_052505} within v2RDM calculations gives one access to reduced density matrices (RDMs) corresponding to the lowest-energy states with a given total spin ($S$). Spin-state splittings obtained from such calculations can be of high quality,\cite{DePrince16_2260} provided that targeted non-singlet states have the maximal spin projection ($M_{S}$). On the other hand, non-maximal $M_{S}$ states generally exhibit energies that are too low,\cite{Ayers12_014110} which implies a troubling qualitative failure of v2RDM theory: for a fixed $S$, the manifold of $M_{S}$ states are not degenerate. Similarly, more recent work\cite{DePrince19_032509} has demonstrated that orbital angular momentum constraints can be enforced to gain access to the lowest-energy orbital angular momentum ($L$) or projection of orbital angular momentum ($M_{L}$) states, in systems that possess these symmetries ({\em i.e.}, atoms and linear molecules, respectively). Unfortunately, similar deficiencies appear in such calculations as have been observed in the spin case. For atomic states with fixed $L$, v2RDM theory yields a set of $M_{L}$ states that are not degenerate, and the best (highest-energy) solution is obtained for maximal $M_{L}$.

This work generalizes the results of Ref.~\citenum{Ayers12_014110} and \citenum{DePrince19_032509} by developing and enforcing constraints on the total angular momentum ($J$) and projection of total angular momentum ($M_{J}$) in atomic systems. Illustrative calculations reveal the following serious deficiencies in v2RDM-based descriptions of $J$ and $M_{J}$ states at the non-relativistic limit ({\em i.e.}, in the absence of spin-orbit coupling): (i) for fixed $S$ and $L$, v2RDM fails to recover the expected degeneracies in different $J$ states, and (ii) for fixed $L$, $S$, and $J$, different $M_{J}$ states are not necessarily degenerate. In addition to these failures, we identify a state for which optimized RDMs are exactly ensemble $N$-representable, and, yet, the energy associated with these RDMs is more than 0.1 E$_{\rm h}$ lower than the exact energy obtained from a configuration interaction (CI) calculation. In this case, we are able to deduce that the 2RDM derives from a linear combination of pure-state 2RDMs and is thus not pure-state $N$-representable. Indeed, we will also demonstrate that 2RDMs optimized for most of the states we consider clearly violate pure-state $N$-representability conditions, even in cases where the energy error relative to CI is small. 

This paper is organized as follows. Section \ref{SEC:THEORY} outlines general aspects of v2RDM theory and describes the total angular momentum and angular momentum projection constraints that we consider. Section \ref{SEC:COMPUTATIONAL_DETAIL} then provides some of the technical details of the calculations we have performed, the results of which are discussed in Sec.~\ref{SEC:RESULTS}. Some concluding remarks can be found in Sec.~\ref{SEC:CONCLUSIONS}. Lastly, the Appendix provides a derivation of two classes of pure-state conditions on the 2RDM.

\section{Theory}
\label{SEC:THEORY}

In this Section, we discuss ensemble $N$-representability conditions that are applied in typical non-relativistic v2RDM calculations, and we introduce additional constraints that should be satisfied for states for which $J$ and $M_J$ are good quantum numbers. Throughout, the labels $p$, $q$, $r$, and $s$ refer to orthonormal spatial orbitals, and Greek labels ($\kappa$, $\lambda$, $\mu$, $\sigma$, and $\tau$) indicate either $\alpha$ or $\beta$ spin.

\subsection{The direct variational determination of the 2RDM}

The electronic energy of a many-electron system is a linear functional of the one-electron reduced density matrix (1RDM) and the 2RDM, the elements of which are 
\begin{equation}
\label{EQN:1RDM}
    {}^1D^{p_{\sigma}}_{q_{\tau}} = \langle \Psi | \hat{a}^\dagger_{p_{\sigma}}\hat{a}_{q_{\tau}} | \Psi \rangle,
\end{equation}
and
\begin{equation}
\label{EQN:2RDM}
    {}^2D^{p_{\sigma}q_{\tau}}_{r_{\kappa}s_{\lambda}} = \langle \Psi | \hat{a}^\dagger_{p_{\sigma}}\hat{a}^\dagger_{q_{\tau}} \hat{a}_{s_{\lambda}}\hat{a}_{r_{\kappa}}| \Psi \rangle,
\end{equation}
respectively.  Here $\hat{a}^\dagger$ and $\hat{a}$ represent fermionic creation and annihilation operators.
For a non-relativistic Hamiltonian, the electronic energy is expressible in terms of the spin-conserving elements of these RDMs as
\small
\begin{align}
    \label{EQN:energy}
        E & =   \frac{1}{2} \sum_{pqrs} ~({}^2D^{p_{\alpha}q_{\alpha}}_{r_{\alpha}s_{\alpha}}
            + {}^2D^{p_{\alpha}q_{\beta}}_{r_{\alpha}s_{\beta}}
            + {}^2D^{p_{\beta}q_{\alpha}}_{r_{\beta}s_{\alpha}}
            + {}^2D^{p_{\beta}q_{\beta}}_{r_{\beta}s_{\beta}} ) ( pr | qs )  \nonumber \\
        &  +  \sum_{pq} ~( {}^1D^{p_{\alpha}}_{q_\alpha} + {}^1D^{p_{\beta}}_{q_\beta} )  h_{pq}
\end{align}
\normalsize
In Eq.~\ref{EQN:energy}, $(pr|qs)$ represents a two-electron repulsion integral \textcolor{black}{in chemists' notation}, $h_{pq}$ represents the sum of electronic kinetic energy and electron-nuclear potential energy integrals, and the summation labels run over all spatial orbitals. The 1RDM and 2RDM can be determined via the minimization of Eq.~\ref{EQN:energy} with respect to variations in their elements. Such a procedure results in physically meaningful RDMs if and only if we can ensure that the RDMs are derivable from a single $N$-electron density matrix (for pure states) or an ensemble of $N$-electron density matrices (for an ensemble state). However, neither pure-state nor ensemble-state $N$-representability are exactly achievable, in general, and only a subset of necessary ensemble $N$-representability constraints is enforced, in practice. 

Trivial $N$-representability constraints include the Hermiticity of the RDMs and their antisymmetry with respect to particle exchange. Slightly more complex constraints can be derived by considering various symmetries that the wave functions from which the RDMs derive should possess. For example, the non-relativistic Hamitonian is particle conserving, so its eigenfunctions should also satisfy
\begin{equation}
\label{EQN:PARTICLE_NUMBER}
\hat{N}| \Psi \rangle = N| \Psi \rangle
\end{equation}
where the particle-number operator, $\hat{N}$ is given by
\begin{equation}
    \hat{N} = \sum_{p}\sum_{\sigma} \hat{a}^\dagger_{p_\sigma}\hat{a}_{p_\sigma}.
\end{equation}
Equation \ref{EQN:PARTICLE_NUMBER} implies a large number of constraints on the 1RDM and 2RDM, the simplest of which being
\begin{equation}
     \label{EQN:TRACE}
     \langle \Psi |\hat{N}| \Psi \rangle = N,
\end{equation}
and
\begin{equation}
     \label{EQN:TRACE2}
     \langle \Psi |\hat{N}^2| \Psi \rangle = N^2.
\end{equation}
The left-hand sides of Eqs.~\ref{EQN:TRACE} and \ref{EQN:TRACE2} are expressible as traces over the 1RDM and the 2RDM, respectively. 
Particle-number symmetry can also be used to generate constraints that connect RDMs of different ranks. For example, one can project Eq.~\ref{EQN:PARTICLE_NUMBER} onto all subspaces defined by $\langle \Psi | \hat{a}^\dagger_{p_\sigma}\hat{a}_{q_\tau}$ as 
\begin{align}
    \label{EQN:CONTRACT_D2_1}
         \forall p_{\sigma}, q_{\tau}:  
         \langle \Psi|\hat{a}^\dagger_{p_{\sigma}}\hat{a}_{q_{\tau}}\hat{N}| \Psi \rangle =  N\langle \Psi | \hat{a}^\dagger_{p_{\sigma}}\hat{a}_{q_{\tau}} |\Psi \rangle.
\end{align}
to obtain contraction relationships between the 2RDM and the 1RDM. 

Additional $N$-representability constraints derive  from the fact that the 1RDM and the 2RDM must have non-negative eigenvalues. Indeed, one can define a hierarchy of constraints defined by the non-negativity of various $p$-body RDMs.\cite{Erdahl01_042113} For example, at the two-particle level ($p=2$), the two-particle RDM (the 2RDM), the two-hole RDM (${}^2${\bf Q}), and the particle-hole RDM (${}^2${\bf G}) should be positive semidefinite, and their non-negativity  constitutes the PQG constraints originally described by Garrod and Percus.\cite{Percus64_1756} In this work, we consider the PQG constraints, as well as a partial three-particle condition known as the T2 constraint.\cite{Erdahl78_697,Percus04_2095}

\subsection{Angular momentum constraints}

{\color{black}For a many-electron system described by a non-relativistic Hamiltonian ($\hat{H}$), one can define simultaneous eigenfunctions of $\hat{H}$, the spin squared ($\hat{S}^2$) operator, and the $z$-projection of spin ($\hat{S}_z$) operator. If we restrict our considerations to atomic many-electron systems, then one can also require that these functions are eigenfunctions of the orbital angular momentum squared ($\hat{L}^2$) and $z$-projection of orbital angular momentum ($\hat{L}_z$) operators. } As such, constraints related to the  spin ($S$){\color{black},} projection of spin ($M_S$){\color{black},  orbital angular momentum ($L$), and projection of orbital angular momentum ($M_L$)} quantum numbers can be derived from the relevant eigenvalue equations,\cite{Ayers12_014110,DePrince19_032509} in a similar manner to what was done above for the case of particle-number symmetry. {\color{black}Alternatively, one could require the wave function to be a simultaneous eigenfunction of $\hat{H}$, $\hat{S}^2$, $\hat{L}^2$, the total angular momentum operator ($\hat{J}^2$), and the $z$-projection of the total angular momentum operator ($\hat{J}_z$). In this case, constraints related to the  total angular momentum ($J$) and projection of total angular momentum ($M_J$) quantum numbers can be similarly derived, but constraints on $M_S$ and $M_L$ can no longer be enforced.} The remainder of this subsection considers constraints on the 1RDM and 2RDM that can be derived by considering wave functions that are simultaneous eigenfunctions of $\hat{S}^2$, $\hat{L}^2$, $\hat{J}^2$, and $\hat{J}_z$. {\color{black} It is worth stressing that the constraints we derive and enforce are expectation value constraints, which are necessary yet insufficient for ensuring that the RDMs derive from eigenfunctions of these operators.}

We first consider the expectation value of the spin squared operator,
\begin{align}                                                                             \label{EQN:S2}
        \langle \Psi |\hat{S}^2| \Psi \rangle &= S(S+1),
\end{align}     
which can be expressed in terms of the 2RDM as
\begin{align}
        \label{EQN:s2}
        &S(S+1) -\frac{3}{4}N\nonumber \\
        = & \sum_{pq}[{}^2D^{p_{\alpha}q_{\beta}}_{q_{\alpha}p_{\beta}} +
        \frac{1}{4}({}^2D^{p_{\alpha}q_{\alpha}}_{p_{\alpha}q_{\alpha}}+
        {}^2D^{p_{\beta}q_{\beta}}_{p_{\beta}q_{\beta}}-
        {}^2D^{p_{\alpha}q_{\beta}}_{p_{\alpha}q_{\beta}}-
        {}^2D^{q_{\alpha}p_{\beta}}_{q_{\alpha}p_{\beta}})]
\end{align}
Here, we have assumed that the 1RDM  satisfies Eq.~\ref{EQN:TRACE}. We also note that this expression is slightly more complicated than that provided in Ref.~\citenum{Mazziotti05_052505} because that work considers RDMs that derive from eigenfunctions of $\hat{S}_z$. No such simplifications can be made for spin-orbit-coupled states, which is the case considered here.

Orbital angular momentum constraints can be derived in a similar way, as described in Ref.~\citenum{DePrince19_032509}. For example, eigenfunctions of $\hat{L}^2$ should satisfy
\begin{equation}
    \label{EQN:L2}
    \langle \Psi | \hat{L}^2 | \Psi \rangle = L(L+1)
\end{equation}
which can be expressed in terms of the elements of the
1RDM and 2RDM as
\begin{equation}
    \label{EQN:L2RDM}
    \sum_{\xi=x,y,z} \bigg ( \sum_{\sigma \tau}\sum_{pqrs} {}^2D^{p_{\sigma}r_{\tau}}_{q_{\sigma}s_{\tau}} [L_\xi]^p_q [L_\xi]^r_s
    + \sum_{\sigma} \sum_{pq} {}^1D^{p_{\sigma}}_{q_{\sigma}} [L_\xi^2]^p_q \bigg )  =  L(L+1).
\end{equation}
Here, $[L_\xi]^p_q$ represents a matrix element of the $\xi$-component of the angular momentum operator $\hat{L}_\xi$, with $\xi \in \{x,y,z\}$. $[L_\xi^2]^p_q$ represents a matrix element of the one-electron component of the square of the $\xi$-component of the angular momentum operator, $\hat{L}_\xi^2$, {\em i.e.}, the second term on the right-hand side of
\begin{equation}
\hat{L}_\xi^2 = \sum_{i \neq j} \hat{L}_\xi(i) \hat{L}_\xi(j)  + \sum_i \hat{L}_\xi(i) \hat{L}_\xi(i),
\end{equation}
where the labels $i$ and $j$ refer to electron coordinates.

The total angular momentum can be expressed as the sum of spin and orbital angular momenta, and, thus, the expectation value of $\hat{J}^2$ is
\begin{align}
    \label{EQN:J2}
    \langle \Psi | \hat{J}^2 | \Psi \rangle &=& \langle \Psi | \hat{L}^2 | \Psi \rangle + \langle \Psi | \hat{S}^2 | \Psi \rangle + 2\langle \Psi | \hat{L}\hat{S} | \Psi \rangle \nonumber \\
    &=& J(J+1).
\end{align}
Assuming that Eqs.~\ref{EQN:S2} and \ref{EQN:L2} are satisfied, a constraint on the expectation value of $\hat{J}^2$ reduces to 
\begin{equation}
    \label{EQN:LS}
    \langle \Psi | \hat{L}\hat{S} | \Psi \rangle = \frac{1}{2}[J(J+1)-S(S+1)-L(L+1)]
\end{equation}
with the spin-orbit operator defined as 
\begin{equation}
   \hat{L}\hat{S} = \hat{L}_x\hat{S}_x+\hat{L}_y\hat{S}_y+\hat{L}_z\hat{S}_z.
\end{equation}
As an example, the second-quantized form of $\hat{L}_x\hat{S_x}$ is
\begin{equation}
   \hat{L}_x\hat{S}_x = \frac{1}{2}\sum_{pq}[L_x]^p_q \sum_{\mu} \hat{a}^\dagger_{p_{\mu}}\hat{a}_{q_{\mu}} 
   \sum_r \sum_{\lambda \neq \sigma}\hat{a}^\dagger_{r_{\lambda}}\hat{a}_{r_{\sigma}},
\end{equation}
where $[L_x]^p_q$ is an integral over the $x$-component of the orbital angular momentum operator. The expectation value of $\hat{L}_x\hat{S}_x$ can be expressed in terms of the 1RDM and 2RDM as
\begin{equation}
    \langle \Psi| \hat{L}_x\hat{S}_x | \Psi \rangle = \frac{1}{2}\sum_{pq}[L_x]^p_q \bigg ( \sum_{\lambda \neq \sigma} \bigg [ {}^1D^{p_{\lambda}}_{q_{\sigma}} + \sum_{\mu} \sum_{r} {}^2D^{p_{\mu}r_{\lambda}}_{q_{\mu}r_{\sigma}} \bigg ] \bigg ).
\end{equation}
Expectation values of  $\hat{L}_y\hat{S}_y$ and $\hat{L}_z\hat{S}_z$ can be expressed in terms in 1RDM and 2RDM in an analogous way. 

The expectation value of the $z$-projection of the total angular momentum 
\begin{equation}
    \label{EQN:Jz}
    \langle \Psi | \hat{J}_z | \Psi \rangle = M_J,
\end{equation}
provides the following constraint on the form of the 1RDM:
\begin{equation}
    \label{EQN:JzRDM}
    \sum_{\sigma}\sum_{pq} {}^1D^{p_{\sigma}}_{q_{\sigma}} [L_z]^p_q  + 
    \frac{1}{2}\sum_{p} ({}^1D^{p_{\alpha}}_{p_{\alpha}}-{}^1D^{p_{\beta}}_{p_{\beta}})= M_J.
\end{equation}
An additional important $\hat{J}_z$-based constraint can be derived by forcing the variance of the $z$-projection of the total angular momentum to vanish, {\em i.e.}, 
\begin{equation}
\label{EQN:JZ_VARIANCE}
\langle \Psi | \hat{J}_z^2|\Psi\rangle -  \langle \Psi | \hat{J}_z|\Psi\rangle^2 = 0
\end{equation}
Assuming Eq.~\ref{EQN:Jz} is satisfied, Eq.~\ref{EQN:JZ_VARIANCE} can be expressed in terms of the 1RDM and the 2RDM as
\begin{align}
    \label{EQN:JZVARRDM2}
    M_J^2 &= 
    \sum_{\sigma \tau}\sum_{pqrs} {}^2D^{p_{\sigma}r_{\tau}}_{q_{\sigma}s_{\tau}} [L_z]^p_q [L_z]^r_s
    + \sum_{\sigma} \sum_{pq} {}^1D^{p_{\sigma}}_{q_{\sigma}} [L_z^2]^p_q  \nonumber \\
    & +\sum_{pq}[L_z]^p_q(\sum_\sigma X_{\sigma \alpha}{}^1D^{p_{\sigma}}_{q_{\sigma}}
    +\sum_{\sigma \tau}\sum_{r} X_{\tau \alpha}
   {}^2D^{p_{\sigma}r_{\tau}}_{q_{\sigma}r_{\tau}})  \nonumber \\
       &+ \frac{1}{4}(\sum_{\sigma \tau} X_{\sigma\tau}\sum_{pq} 
       {}^2D^{p_{\sigma}q_{\tau}}_{p_{\sigma}q_{\tau}}+\sum_{\sigma} \sum_{pq} {}^1D^{p_{\sigma}}_{q_{\sigma}}),
\end{align}
where the $X_{\sigma \tau}$ is a sign function defined as
\begin{equation}
    X_{\sigma \tau} :=
\begin{cases}
    1, & \text{if } \sigma = \tau\\
    -1,& \text{if } \sigma \neq \tau
\end{cases}.
\end{equation}
We have previously shown \cite{DePrince19_032509} that a variance constraint similar to this one can dramatically improve the quality of v2RDM-based descriptions of states with non-maximal $z$-projections of the orbital angular momentum. 

Lastly, we note the existence of classes of constraints defined by the raising and lowering operators
\begin{equation}
    \hat{J}_\pm = \hat{J}_x \pm i\hat{J}_y
\end{equation}
For example, maximal eigenfunctions of $\hat{J}_z$ should satisfy $J_+|\Psi_{M_J=J} \rangle = 0$ and $J_-|\Psi_{M_J=-J} \rangle = 0$, and the former equality can be projected onto all subspaces defined by $\langle \Psi_{M_J=J} | \hat{a}^\dagger_{p_\sigma}\hat{a}_{q_\tau}$ to yield
\begin{equation}
\label{EQN:J+}
 \forall p_{\sigma}, q_{\tau}:  
   \langle \Psi_{M_J=J} | \hat{a}^\dagger_{p_\sigma}\hat{a}_{q_\tau}\hat{J}_+|\Psi_{M_J=J} \rangle = 0
\end{equation}
Like all other constraints considered herein, Eq.~\ref{EQN:J+} is expressible in terms of the elements of the 1RDM and 2RDM. Maximal $z$-projection constraints relevant to spin\cite{Ayers12_014110} and orbital\cite{DePrince19_032509} angular momentum have  previously been shown to be weak or effectively inactive constraints on the RDMs, and numerical tests suggest that Eq.~\ref{EQN:J+} also appears to be inactive when Eq.~\ref{EQN:JZ_VARIANCE} is satisfied. Prior work has demonstrated that raising/lowering constraints involving non-maximal spin-projection states are active\cite{Ayers12_014110} but ineffective at restoring expected degeneracies in $M_S$ states. In this work, we do not consider such raising/lowering constraints on non-maximal $M_J$ states.

\section{Computational Details}
\label{SEC:COMPUTATIONAL_DETAIL}

The calculations discussed herein considered electronic states in second-row atoms (Be, B, C, and N) that can be generated by excitations within the valence orbitals (no Rydberg states were considered). Reference energies were taken from non-relativistic full CI calculations carried out using the \textsc{Psi4} quantum chemistry package.\cite{Sherrill20_184108} The v2RDM approach was implemented as a plugin to \textsc{Psi4}, and all v2RDM calculations used a non-relativistic Hamiltonian, with all orbitals correlated (unless otherwise noted). All v2RDM computations exploited the block structure of the RDMs resulting from abelian point-group symmetry, but the point group was chosen such that all operators belonged to the totally symmetric irreducible representation. When considering angular momentum operators, the largest point group that satisfies this requirement is $C_i$. {\color{black}As a result, the present calculations are more computationally demanding than those for atomic systems in the absence of total angular momentum constraints.} {\color{black}We consider v2RDM calculations enforcing either two-particle (PQG) or PQG plus partial three-particle (T2) $N$-representability conditions. Under these conditions, the computational cost of these calculations increases with the sixth or ninth power with the number of active orbitals, respectively. This scaling is unchanged by the introduction of angular momentum constraints.} All full CI and v2RDM calculations were carried out within the STO-3G\cite{Pople69_2657} and the {\color{black}6-31G} basis sets.\cite{Dunning89_1007} Unless otherwise noted, v2RDM calculations were considered converged when the primal-dual energy gap fell below $1\times10^{-4}$ E$_{\rm h}$ and the primal and dual constraint errors fell below $1\times10^{-6}$. For details regarding the definitions of these quantities, the reader is referred to Ref.~\citenum{DePrince19_6164}.

\section{Results and Discussion}
\label{SEC:RESULTS}

\subsection{Non-degeneracy of total angular momentum states}
\label{SUBSEC:DEGENERACY}
\begin{figure}
    \caption{Errors in v2RDM-derived energies (10$^{-3}$ E$_{\rm h}$) for a carbon atom in (a)$^3$P$_0$, $^3$P$_1$, and $^3$P$_2$ (b) $^3$D$_1$, $^3$D$_2$, and $^3$D$_3$ states with different $z$-projections of the total angular momentum.$^a$ {\color{black} An error of zero corresponds to exact agreement with the full CI reference value. All calculations were performed within the STO-3G basis set.} }
    \label{FIG:C_3P_3D_ENERGY}
    \begin{center}
        \includegraphics[scale=1.0]{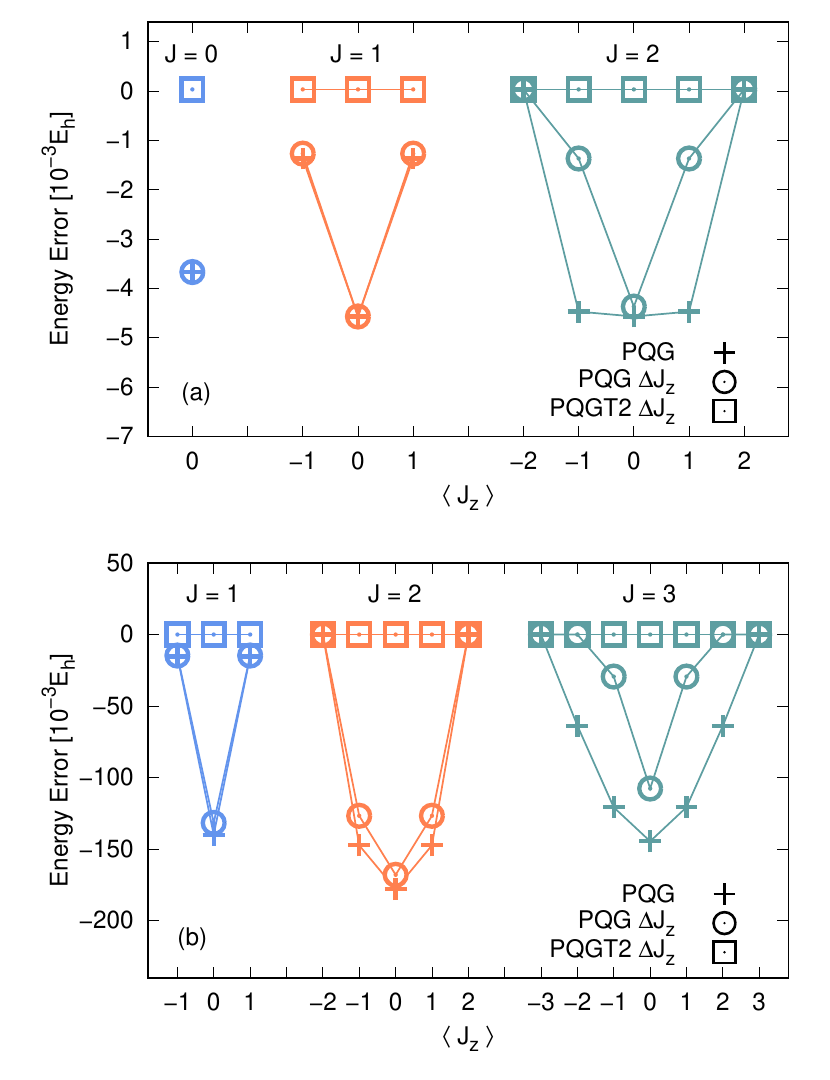}
    \end{center}
    {\scriptsize ${}^a$ The primal-dual energy gaps for PQG $\Delta$J$_z$ calculations on the $M_J = 1$ projections of the ${}^3$D$_1$ and $^{}3$D$_3$ states were converged to $3 \times 10^{-4}$ E$_{\rm h}$. The energy gap for the PQG calculation on the $M_J = 1$ projection of the ${}^3$D$_1$ state was converged to $4 \times 10^{-4}$ E$_{\rm h}$.}
\end{figure}

We begin this Section by establishing that v2RDM calculations that consider incomplete ensemble $N$-representability conditions do not provide a balanced description of what should be energetically equivalent total angular momentum states. Figure \ref{FIG:C_3P_3D_ENERGY} depicts errors in v2RDM-derived energies for the $^3$P and $^3$D states of the carbon atom in a minimal basis set, with full CI energies serving as the reference. Several sets of data are provided. The label ``PQG'' indicates that the optimized RDMs satisfy standard two-body $N$-representability conditions, plus constraints on the expectation values of $\hat{L}^2$, $\hat{S}^2$, $\hat{J}^2$, and $\hat{J}_z$. The label ``PQG $\Delta J_z$'' indicates the addition of the variance constraint given in Eq.~\ref{EQN:JZVARRDM2}. Lastly, the label ``PQGT2 $\Delta J_z$'' denotes that the optimized RDMs satisfy the PQG and T2 conditions, as well as expectation-value- and variance-based angular momentum constraints. 
We find that calculations in which RDMs satisfy the PQG and T2 conditions recover the correct degeneracies in $J$ and $M_J$ states for the cases considered in Fig.~\ref{FIG:C_3P_3D_ENERGY}. However, RDMs that satisfy only the PQG conditions clearly do not, and, in several cases, absolute energy errors from PQG calculations exceed 0.1 E$_{\rm h}$. 

We can make the following general statements regarding the behavior of v2RDM calculations performed under the PQG constraints. For a given $J$, maximal $M_J$ states are the best constrained, meaning they have the highest energy.
In some cases, constraints on the variance of $\hat{J}_z$ improve the description of the non-maximal angular momentum projection states, but such constraints do not improve the description of the maximal projection states, nor do they restore the expected degeneracy of different angular momentum projections. This behavior is consistent with that observed for orbital and spin angular momentum projection states (see Refs.~\citenum{DePrince19_032509} and \citenum{Ayers12_014110}, respectively). We also note that PQG calculations fail to recover the expected degeneracy between different $J$ states corresponding to the same $L$ and $S$, even for the maximal projection cases. For states with $|M_J|$ = $J$, the best (highest-energy) results are obtained for the highest $J$.

\begin{figure*}
    \caption{Errors in v2RDM-derived energies (10$^{-3}$ E$_{\rm h}$) for   different total angular momentum projections of the $^2$D$_{3/2}$ and $^2$D$_{5/2}$ states of a nitrogen atom. {\color{black}Results are provided for optimized RDMs that satisfy the (a) PQG or (b) PQG+T2 constraints. An error of zero corresponds to exact agreement with full CI reference value.  Results are provided for calculations in the STO-3G and 6-31G basis sets$^a$ in which all electrons and orbitals were correlated.}
    }
    \label{FIG:F_2P_BASIS_SET_EFFECT}
    \begin{center}
        \includegraphics[scale=1.0]{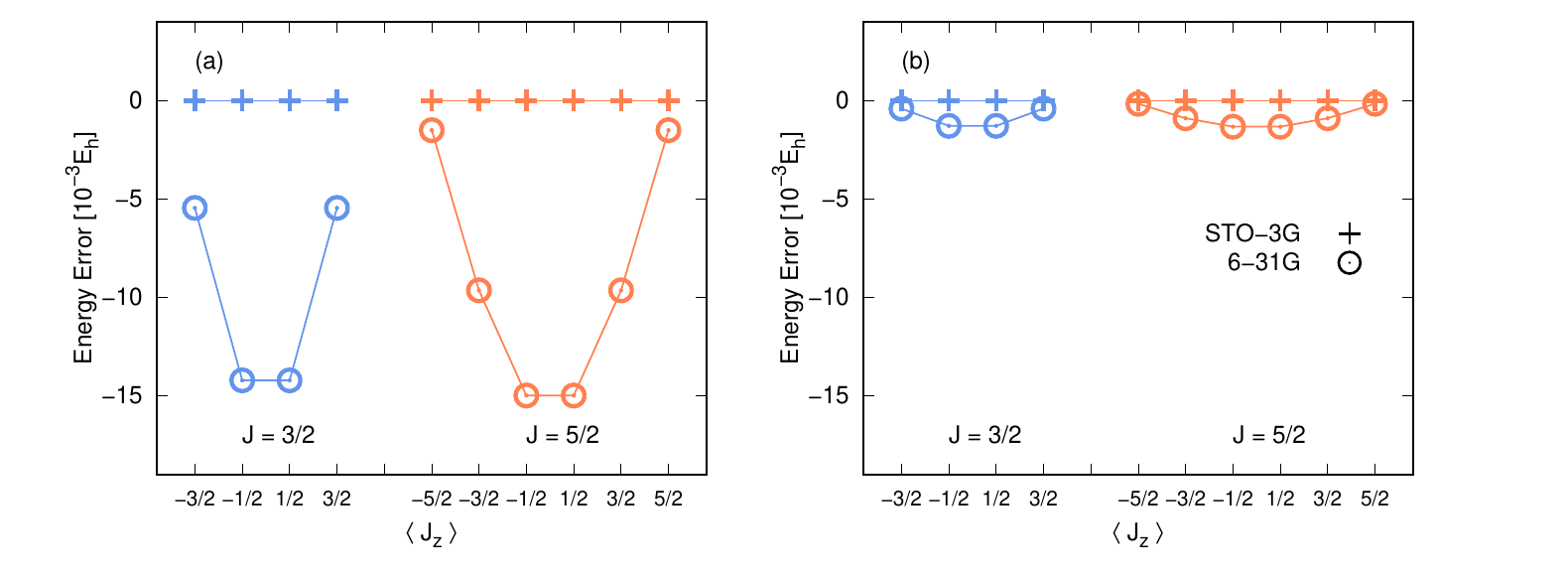}
    \end{center}
    {\scriptsize {\color{black}$^a$ With the exception of the $M_J$ = $\pm $1/2 projections of the ${}^2$D$_{3/2}$ state, PQG+T2 calculations in the 6-31G basis set were converged to a primal-dual energy gap of less than 10$^{-4}$ E$_{\rm h}$ and primal and dual errors of less than 10$^{-5}$. The PQG+T2/6-31G calculations describing the $M_J$ = $\pm$ 1/2 projections of the ${}^2$D$_{3/2}$ state used convergence thresholds of 2 $\times$ 10$^{-4}$ E$_{\rm h}$ and $10^{-5}$, respectively. }}
\end{figure*}

Figure \ref{FIG:F_2P_BASIS_SET_EFFECT} examines the basis set dependence of the degeneracy issues uncovered above. Here, we consider errors in v2RDM-derived energies for the ${}^2$D$_{3/2}$ and ${}^2$D$_{5/2}$ states of a nitrogen atom, with full CI energies serving as the reference. The v2RDM-optimized RDMs satisfy PQG {\color{black}or PQG+T2} $N$-representability conditions, as well as the expectation-value- and variance-based constraints discussed above. Two basis sets are considered. In a minimal (STO-3G) basis, v2RDM correctly captures the expected degeneracy of all $J$ and $M_J$ states {\color{black} when PQG or PQG+T2 constraints are enforced}. However, in the larger ({\color{black}6-31G}) basis, the degeneracy among these states is broken, and the general conclusions regarding the behavior of PQG calculations discussed above apply to these data. {\color{black}Energy errors can be as large as 15 $\times 10^{-3}$ E$_{\rm h}$ when enforcing the PQG constraints. When RDMs satisfy the PQG+T2 constraints, this error is substantially reduced, but the non-degeneracy persists.} This example suggests that the non-degeneracy issue becomes more pronounced when correlating larger numbers of orbitals.

\begin{figure}
    \caption{Errors in v2RDM-derived energies of $^3$P spin and orbital angular momentum projection states of a carbon atom. {\color{black} An error of zero corresponds to exact agreement with full CI reference value. All calculations were performed within the STO-3G basis set.}
    }
    \label{FIG:C_3P_L}
    \begin{center}
        \includegraphics[scale=1.0]{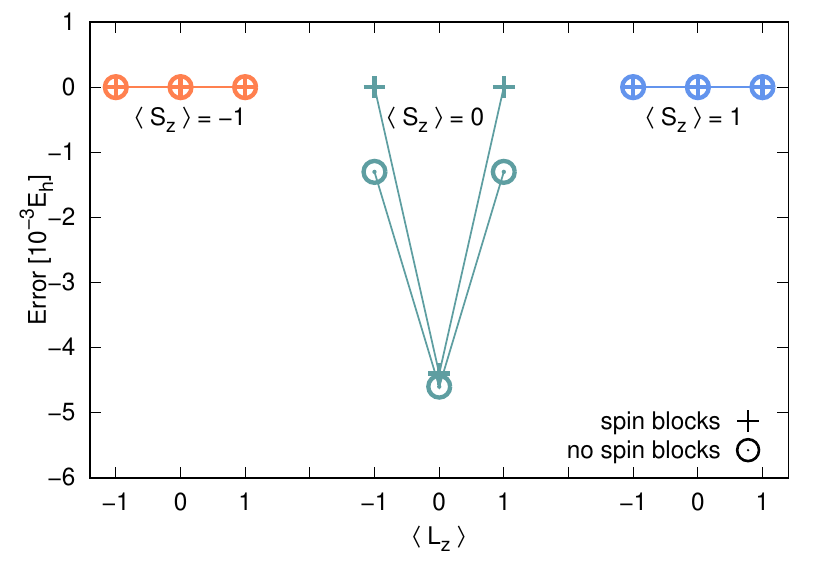}
    \end{center}
\end{figure}

Before moving on, we highlight two subtle yet important differences between v2RDM calculations that place constraints on $S$, $L$, $M_L$, and $M_S$ versus those that place constraints on $S$, $L$, $J$, and $M_J$. We refer to these cases as uncoupled and coupled angular momentum states, respectively. First, for uncoupled states, we preserve the spin-blocked structure expected for RDMs that derive from wave functions having good spin-projection quantum numbers, whereas this block structure is relaxed for coupled states. Figure \ref{FIG:C_3P_L} provides errors in v2RDM-derived energies for ${}^3$P spin and orbital angular momentum projection states of a carbon atom. Here, results labeled ``spin blocks'' and ``no spin blocks'' correspond to calculations performed under the PQG conditions, with constraints on the expectation values of $\hat{L}^2$, $\hat{S}^2$, $\hat{L}_z$, and $\hat{S}_z$, as well as constraints on the variances of $\hat{L}_z$ and $\hat{S}_z$; the only difference between the calculations is the block structure imposed on the RDMs. We can see that, for maximal spin projections ($M_S$ = $\pm 1$), different orbital angular momentum projection states are degenerate, regardless of the block structure of the RDMs. On the other hand, both sets of calculations fail to retain the correct degeneracy in the orbital angular momentum projection states when $M_S = 0$, and worse results are obtained for the case that relaxes the spin block structure. Hence, we can conclude that the additional variational flexibility introduced via the relaxed spin block structure in coupled angular momentum calculations exaggerates the existing  degeneracy issues in the uncoupled case.

\begin{figure}
    \caption{Errors in v2RDM-derived energies for various $^2$P angular momentum states of a boron atom.  {\color{black}An error of zero corresponds to exact agreement with full CI reference value. All calculations were performed within the STO-3G basis set.}}
    \label{FIG:B_2P_CG}
    \begin{center}
        \includegraphics[scale=1.0]{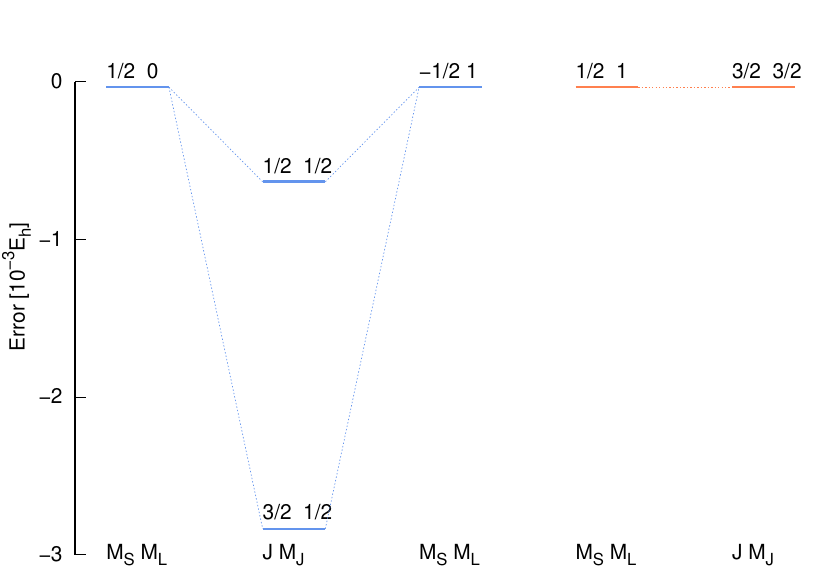}
    \end{center}
\end{figure}

The second difference between v2RDM calculations describing uncoupled and coupled angular momentum states is that the states themselves are different; coupled angular momentum states are admixtures of uncoupled states, {\color{black}according to Clebsch--Gordan coefficients}. Figure \ref{FIG:B_2P_CG} illustrates errors in v2RDM-derived energies for the ${}^2$P states of a boron atom, and the schematic connects coupled angular momentum states to the uncoupled states to which they should correspond. Here, calculations on uncoupled and coupled states both consider RDMs that satisfy the PQG conditions. Uncoupled calculations consider constraints on $S^2$, $L^2$, $M_S$, and $M_L$, while coupled calculations consider constraints on $S^2$, $L^2$, $J^2$, and $M_J$. Neither set of calculations imposes spin-block structure on the RDMs so that we can focus on the angular momentum constraints themselves. In this case, we find that the energies of uncoupled states are well-described by v2RDM theory and that the approached exhibits no degeneracy issues. On the other hand, coupled states are not degenerate, and  the correct energy is only obtained for the maximal $J$ state with $M_J = J$. The data in Fig.~\ref{FIG:B_2P_CG} suggest that the energies of coupled angular momentum states are a lower bound to those of uncoupled states, but this is not actually the case, in general. A similar analysis for the $^3$P state of carbon can be found in the Supporting Information; in this case, some coupled states lie higher in energy than the parent uncoupled states.

\subsection{Ensemble- versus pure-state $N$-representability}
\label{SEC:GPC}

All v2RDM-derived results presented in this work were generated under ensemble $N$-representability conditions. In this Section, we draw connections between the degeneracy issues discussed above and violations in pure-state conditions not enforced in our calculations. A useful case study for this purpose is the ${}^1$P$_1$ state of atomic beryllium. Moving forward, for v2RDM calculations on this system and for all remaining v2RDM optimizations, we enforce constraints on the expectation values of $\hat{S}^2$, $\hat{L}^2$, $\hat{J}^2$, and $\hat{J}_z$, as well as on the variance of $\hat{J}_z$. Table \ref{TABLE:BE_1P1} provides errors in energies from v2RDM calculations where RDMs satisfied the PQG or PQG plus T2 constraints, as compared to results taken from full CI calculations. While PQG constraints lead to accurate energies for the maximal projections ($M_J = \pm 1$), these constraints do not provide a good description of the $M_J = 0$ state. In this case, v2RDM energies are 0.1679 E$_{\rm h}$ too low. Moreover, the addition of the T2 constraint only improves the energy by 0.0001 E$_{\rm h}$. This failure comes as a surprise, as Be only has two valence electrons, so one would expect that PQG constraints should provide an accurate description of its electronic structure, as in the case of the maximal projection states tabulated in Table \ref{TABLE:BE_1P1}.

\begin{table}[]
\caption{Errors in v2RDM-derived energies (10$^{-3}$ E$_{\rm h}$) for different total angular momentum projections of the ${}^1$P$_1$ state of beryllium.}
\label{TABLE:BE_1P1}
\begin{tabular}{cccc}
\hline\hline
$M_J$ &~& \multicolumn{1}{c}{PQG}  & \multicolumn{1}{c}{PQGT2} \\
\hline
-1 &~&     0.0  &    0.0  \\
0  &~&  -167.9  & -167.8  \\
1  &~&     0.0  &    0.0  \\
\hline\hline             
\end{tabular}
\end{table}

\subsubsection{Generalized Pauli constraints for the $^{1}$P$_1$ state of beryllium}

While necessary ensemble $N$-representability conditions on $p$-body RDMs are straightforward to devise,\cite{Erdahl01_042113} and a complete solution to the ensemble $N$-representability problem has been put forward,\cite{Mazziotti12_263002} the pure-state $N$-representability problem is much more complex. Indeed, until only a short while ago,\cite{Klyachko06_72,Klyachko08_287} necessary and sufficient pure-state conditions on the 1RDM, known as generalized Pauli constraints (GPCs), were only known for special cases.\cite{Dennis72_7} 
In Ref.~\citenum{Klyachko08_287}, however, Altunbulak and Klyachko developed an algorithm to identify GPCs for general systems and tabulated these constraints for systems involving up to ten spin orbitals. 
{\color{black}These GPCs take the form of inequality constraints on the eigenvalues of the 1RDM (the natural orbital occupation numbers). As an example, consider the following inequality constraint that applies to the natural orbital occupation numbers for a system of three electrons distributed among six spin orbitals:
\begin{equation}
    \lambda_5 + \lambda_6 - \lambda_4 \ge 0
\end{equation}
Here, $\lambda_4$, $\lambda_5$, and $\lambda_6$ represent the occupations of the fourth, fifth, and sixth natural orbitals, when they are ordered with decreasing occupations. If the left-hand side of this equation evaluated to -0.1, then we would say that this condition is violated by -0.1 electrons.}
The GPCs also indirectly apply to the 2RDM.\cite{Mazziotti16_032516} As shown in the Appendix, these constraints can be applied to a 1RDM for any $(N-1)$- or $(N+1)$-electron systems generated by the removal or addition of an electron from or to the $N$-electron state, respectively; such constraints can be expressed in terms of the 2RDM for the $N$-electron state. Enforcing GPCs can be challenging, in practice.\cite{DePrince16_164109, Helbig15_154108} In this work, we merely assess whether 1RDMs and 2RDMs optimized under ensemble $N$-representability constraints satisfy these conditions. 

We begin with the troubling case of the ${}^1$P$_1$ state of beryllium, as described by v2RDM calculations that enforce the PQG constraints, within STO-3G basis set. The family of antisymmetrized wave functions corresponding to this four electron / ten spin orbital system is represented by the symbol  $\wedge^4\mathcal{H}_{10}$. 
Table \ref{TABLE:Be_FULL_GPC} provides the energy error and GPC errors for the $M_J = 0$ and $M_J = 1$ projections of the ${}^1$P$_1$ state. The {\color{black}GPC} errors are defined as the sum of all {\color{black}observed violations in the GPCs}, and a negative value indicates that {\color{black} at least one condition was violated; in the three-electron-in-six-spin-orbital example above, the tabulated GPC error would have been -0.1 electrons.} The notation $N-1$ and $N+1$ refer to GPCs applied to the 1RDM for the $(N-1)$- or $(N+1)$-electron states, respectively,
and wave functions for these systems belong to the $\wedge^3\mathcal{H}_{10}$ and $\wedge^5\mathcal{H}_{10}$ families, respectively. 
We note that the $(N \pm 1)$-electron wave functions are defined by the expansions given in the Appendix (see Eqs.~\ref{EQN:NM1_STATE} and \ref{EQN:NP1_STATE}, respectively), with the expansion coefficients assumed to be real and chosen to maximize the magnitude of the errors in the GPCs (see the Appendix for a more detailed description of this procedure). As mentioned above, such conditions actually apply to the 2RDM for the $N$-electron state. From these data, we first note that the GPCs on the 1RDM for the $N$-electron state are satisfied. However, GPCs applied to the 2RDM are violated for both projections, with the larger errors observed for the $M_J = 0$ case (the case for which the energy is poorly described by ensemble conditions). Hence, it appears that large energy errors are associated with violations in GPCs on the 2RDM, but the converse cannot be stated. {\color{black}As is shown below, a} small energy error does not guarantee that the GPCs on the 2RDM are necessarily satisfied. 

\begin{table}[]
\centering
\caption{Energy errors (10$^{-3}$ E$_{\rm h}$) and GPC errors {\color{black}(in units of electrons)} for the $M_J = 0$ and $M_J = 1$ projections of the ${}^1$P$_1$ state of beryllium. The full CI and v2RDM calculations correlated all four electrons in all ten spin orbitals. }
\label{TABLE:Be_FULL_GPC}
\begin{tabular}{cccccc}
\hline
\hline
 & & & \multicolumn{3}{c}{GPC errors} \\
 \cline{4-6}
 & & &      & \multicolumn{2}{c}{2RDM} \\
 \cline{5-6}
 & $M_J$ & \multicolumn{1}{c}{energy error} & \multicolumn{1}{c}{1RDM} & \multicolumn{1}{c}{$N-1$} & \multicolumn{1}{c}{$N+1$} \\
\hline
${}^1$P$_1$ & 0 & -167.9 & 0.00 & -350.86 & -91.34 \\
${}^1$P$_1$ & 1 & 0.0 & 0.00 & -22.62 & 0.00 \\
                               \hline\hline
\end{tabular}
\end{table}

Given the large violations in the GPCs on the 2RDM tabulated in Table \ref{TABLE:Be_FULL_GPC}, it is interesting to consider the same states, but for the case where we do not correlate the electrons in the 1s orbital. This situation corresponds to an active-space v2RDM calculation on two electrons in eight spin orbitals, and the PQG constraints are complete ensemble $N$-representability conditions in this case. Table \ref{TABLE:Be_GPC} provides the energy error and GPC errors for the $M_J = 0$ and $M_J = 1$ projections of the ${}^1$P$_1$ state, with the 1s electrons frozen. The results for the $M_J = 1$ state are unsurprising; the energy error is zero, and the GPCs on the 1RDM and 2RDM are all satisfied. However, the error in the v2RDM energy for the $M_J = 0$ state is almost as large as that for the full space calculation (-0.1675 E$_{\rm h}$, as compared to the energy from an active space full CI calculation with the same active space), despite the fact that the v2RDM-optimized 1RDM and 2RDM are \textit{exactly} ensemble $N$-representable. In this case, we find that GPCs on the 1RDM are satisfied, as are the GPCs on the 1RDM for the ($N-1$)-electron state. On the other hand, the GPCs on the 1RDM for the ($N+1$)-electron state are violated, and we can thus conclude that the energy error is related to the ensemble nature of the 2RDM.

\begin{table}[]
\centering
\caption{Energy errors (10$^{-3}$ E$_{\rm h}$) and GPC errors {\color{black}(in units of electrons)} for the $M_J = 0$ and $M_J = 1$ projections of the ${}^1$P$_1$ state of beryllium. The full CI and v2RDM calculations correlated an active space of two electrons in eight spin orbitals. }
\label{TABLE:Be_GPC}
\begin{tabular}{cccccc}
\hline
\hline
 & & & \multicolumn{3}{c}{GPC errors} \\
 \cline{4-6}
 & & &      & \multicolumn{2}{c}{2RDM} \\
 \cline{5-6}
  & $M_J$ & \multicolumn{1}{c}{energy error} & \multicolumn{1}{c}{1RDM} & \multicolumn{1}{c}{$N-1$} & \multicolumn{1}{c}{$N+1$} \\
\hline
${}^1$P$_1$ & 0 & -167.5 & 0.00 & 0.00 & -22.57 \\
${}^1$P$_1$ & 1 &    0.0 & 0.00 & 0.00 &   0.00 \\
                               \hline\hline
\end{tabular}
\end{table}

\subsubsection{Wave function analysis for the ${}^1$P$_1$ state of beryllium}

With the 1s electrons frozen, the v2RDM-optimized 2RDM for the $M_J = 0$ projection of the $^1$P$_1$ state of beryllium is exactly ensemble $N$-representable but not pure-state $N$-representable. This property implies that the 2RDM is derivable from a sum of at least two pure-state 2RDMs. Interestingly, this example is simple enough for us to deduce numerically a set of two-electron wave functions whose 2RDMs can be combined to yield the ensemble $N$-representable 2RDM we obtain from v2RDM theory.

For this state, we expand all quantities in the basis of real-valued Hartree-Fock orbitals for the ground state (${}^1$S$_0$). The 1s orbital is frozen at double occupation, and the 2s, 2p$_x$, 2p$_y$, and 2p$_z$ orbitals are active.
{\color{black}In this basis, the true wave function for the $M_J = 0$ projection of the $^1$P$_1$ state is simply an open-shell singlet of the form $|\Psi \rangle = \frac{1}{\sqrt{2}} ( | 1s^2 2s(\alpha) 2p_z(\beta) \rangle + | 1s^2 2s(\beta) 2p_z(\alpha) \rangle )$, where $| 1s^2 2s(\alpha) 2p_z(\beta) \rangle$ represents a determinant with a doubly occupied 1s orbital and singly occupied 2s and 2p$_z$ orbitals (occupied by electrons with $\alpha$- and $\beta$-spin, respectively). The associated 1RDM is diagonal, with equivalent spin blocks (${}^1D^\alpha_\alpha = {}^1D^\beta_\beta$), and equal occupations of the 2s and 2p$_z$ orbitals. The 2p$_x$ and 2p$_y$ orbitals have zero occupation. The v2RDM-optimized 1RDM for this state, on the other hand, has a different structure. It is diagonal, with equivalent spin blocks, but the occupations of the 2p$_x$ and 2p$_y$ orbitals are non-zero, and the occupations of the 2s and 2p$_z$ orbitals differ. This structure implies that configurations other than the open-shell singlet contribute to the ensemble that defines the 2RDM from which the 1RDM derives.  Indeed, the simplest ensemble that can be constructed that recovers the v2RDM-optimized 2RDM has}  seniority-zero structure, \textit{i.e.} only determinants with all electrons paired comprise the wave function expansions. {\color{black}Specifically, these wave functions have the form}
\begin{eqnarray}
    \label{EQN:WFN}
    |\Psi \rangle &=& c_0 | 1s^2 2s^2 \rangle + c_1 | 1s^2 2p_x^2 \rangle \nonumber \\
    &+& c_2 | 1s^2 2p_y^2 \rangle+ c_3 | 1s^2 2p_z^2 \rangle
\end{eqnarray}

Numerical analysis indicates that no single wave function of the form of Eq.~\ref{EQN:WFN} can yield the 2RDM obtained from our v2RDM calculations, but two wave functions, $|\Psi_1\rangle$ and $|\Psi_2\rangle$,
can as
\begin{eqnarray}
    {}^2D^{pq}_{rs} &=& 0.588 \langle \Psi_1| \hat{a}^\dagger_p \hat{a}^\dagger_q \hat{a}_s \hat{a}_r |\Psi_1 \rangle \nonumber \\
    &+& 0.412 \langle \Psi_2 |\hat{a}^\dagger_p \hat{a}^\dagger_q \hat{a}_s \hat{a}_r |\Psi_2 \rangle.
\end{eqnarray}
The expansion coefficients that define $|\Psi_1\rangle$ and $|\Psi_2\rangle$ are provided in Table \ref{TABLE:WFN_COEFS}.
What is interesting here is that neither $|\Psi_1\rangle$ nor $|\Psi_2\rangle$ are  eigenfunctions $\hat{L}^2$ (see Table \ref{TABLE:WFN_COEFS}), whereas the expectation value of this operator is constrained within the v2RDM calculation. Indeed, we can see that the ensemble 2RDM generated from these wave functions does have the correct value for $L(L+1)$, \textit{i.e.}, $1.378 \times 0.588 + 2.890 \times 0.412 \approx 2$. Hence, it appears that the low energy from v2RDM theory derives from the fact that the ensemble that defines the 2RDM does not have a good orbital angular momentum quantum number. Perhaps constraints on the variance of $\hat{L}^2$ could rectify this situation, but this quantity depends on the four-particle RDM and would thus be challenging to constrain in practice.

\begin{table}[]
\centering
\caption{Wave function expansion coefficients for two wave functions whose 2RDMs map onto the v2RDM-generated ensemble 2RDM for $M_J = 0$ $^1P_1$ of Be. The expectation value of $\hat{L}^2$ is also provided for these two wave functions.}
\label{TABLE:WFN_COEFS}
\begin{tabular}{ccc}
\hline
\hline
 & \multicolumn{1}{c}{$|\Psi_1 \rangle$} & \multicolumn{1}{c}{$|\Psi_2 \rangle$}\\
 \hline
  $c_0$ & -0.831 & -0.681 \\
  $c_1$ &  0.359 & -0.149 \\
  $c_2$ &  0.359 & -0.149 \\
  $c_3$ & -0.228 &  0.701 \\
  \hline
  $\langle \hat{L}^2 \rangle$ & 1.377 & 2.890 \\

                              \hline\hline
\end{tabular}
\end{table}

\subsubsection{Generalized Pauli constraints for other atomic systems}

Lastly, we consider the relationship between energy errors and GPC errors for v2RDM-optimized total angular momentum states in two additional atomic systems, boron and carbon, as described by v2RDM theory under PQG conditions and within the STO-3G basis set. With all electrons correlated among all orbitals, wave functions for these systems belong to the $\wedge^5\mathcal{H}_{10}$ and $\wedge^6\mathcal{H}_{10}$ families, respectively. Note, however, that we actually apply GPCs to the one-hole RDM of carbon, and that these conditions are equivalent to those of $\wedge^4\mathcal{H}_{10}$ wave functions. For GPCs applied to the 2RDM, we consider here only the ($N-1$)-electron states that belong to the $\wedge^4\mathcal{H}_{10}$ and $\wedge^5\mathcal{H}_{10}$ families in the case of boron and carbon, respectively.

\begin{figure}
    \caption{{\color{black}Energy errors (10$^{-3}$ E$_{\rm h}$) and 2RDM GPC errors (in units of electrons) for v2RDM-optimized total angular momentum states in boron and carbon atoms.}}
    \label{FIG:GPC}
    \begin{center}
        \includegraphics[scale=1.0]{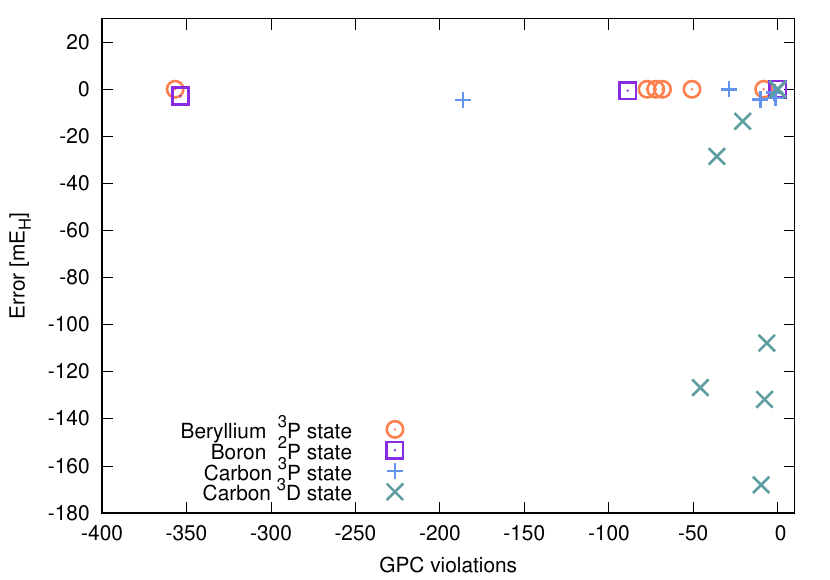}
    \end{center}
\end{figure}

Table \ref{TABLE:GPC} provides energy and GPC errors for v2RDM-optimized total angular momentum states in boron and carbon{\color{black}, as well as some additional states for beryllium that were not considered above}. One notable observation is that the GPCs on the 1RDM are satisfied for all states {\color{black}calculated for boron and carbon}, regardless of the observed energy error. This result stands in stark contrast to those of Ref.~\citenum{DePrince16_164109}, where 1RDMs optimized without any orbital or total angular momentum constraints displayed large errors in GPCs for the ground states of boron (${}^2$P) and carbon (${}^3$P). {\color{black}The only exception is for the $M_J = 0$ projection of the ${}^3$P$_1$ state of beryllium, where we find that the GPCs on the 1RDM are violated, despite there being zero energy error.} 
{\color{black}As for GPCs on the 2RDM, we find that these conditions} are violated for all cases that exhibit non-zero energy error, but {\color{black}the converse cannot be stated: zero energy error does not imply zero error in the GPCs for the 2RDM. Moreover, for states with non-zero energy error}, the magnitudes of the GPC errors do not correlate well with the magnitudes of the energy errors. For example, the absolute energy error of the $M_J = 0$ projection of the ${}^3$D$_2$ state of carbon is 37 times larger than that for the $M_J = 0$ projection of the ${}^3$P$_1$ state, and, yet, the error in the GPCs on the 2RDM is 19 times larger for the latter state. {\color{black}This lack of correlation can clearly be seen in Fig.~\ref{FIG:GPC}, which provides a visual representation of the energy and GPC errors tabulated in Table \ref{TABLE:GPC}.} Lastly, we note that one of the states with zero energy error, the $M_J = 3$ projection of the ${}^3$D$_3$ state of carbon, is a Hartree-Fock state. The optimized 1RDM is idempotent, and the 2RDM is simply an antisymmetrized product of the 1RDM with itself, as expected, and thus both the 1RDM and 2RDM are exactly pure-state $N$-representable.

\begin{table}[]
\centering
\caption{Energy errors (10$^{-3}$ E$_{\rm h}$) and {\color{black} 2RDM} GPC errors {\color{black}(in units of electrons)} for v2RDM-optimized total angular momentum states in boron and carbon atoms.}
\label{TABLE:GPC}
\begin{tabular}{cccccc}
\hline
\hline
& & & & \multicolumn{2}{c}{GPC errors} \\
\cline{5-6}
                              &             & J$_z$  & \multicolumn{1}{c}{energy error} & \multicolumn{1}{c}{1RDM} & \multicolumn{1}{c}{2RDM} \\ 
                              \hline 

 \multirow{6}{*}{\textcolor{black}{Be}}     &  \textcolor{black}{$^3P_{0}$} & 	\textcolor{black}{0} & 	\textcolor{black}{0.0} &   \textcolor{black}{0.00} & \textcolor{black}{-68.22} \\
 & \textcolor{black}{$^3P_{1}$} &     \textcolor{black}{0} & 	\textcolor{black}{0.0}  &  \textcolor{black}{-16.07} & \textcolor{black}{-356.69} \\
 & \textcolor{black}{$^3P_{1}$} & 	\textcolor{black}{1} & 	\textcolor{black}{0.0}  &   \textcolor{black}{0.00} & \textcolor{black}{-72.25} \\
 & \textcolor{black}{$^3P_{2}$} &     \textcolor{black}{0} & 	\textcolor{black}{0.0}  &   \textcolor{black}{0.00} & \textcolor{black}{-77.39} \\
 & \textcolor{black}{$^3P_{2}$} & 	\textcolor{black}{1} & 	\textcolor{black}{0.0}  &   \textcolor{black}{0.00} & \textcolor{black}{-8.29} \\
 & \textcolor{black}{$^3P_{2}$} & 	\textcolor{black}{2} & 	\textcolor{black}{0.0}  &   \textcolor{black}{0.00} & \textcolor{black}{-50.79} \\
 \hline
 \multirow{3}{*}{B}     &  $^2P_{1/2}$ & 	1/2 & 	-0.6 &   0.00 & -188.78 \\
  & $^2P_{3/2}$	 & 1/2 & 	-2.9	 &   0.00 & -353.55 \\
 & $^2P_{3/2}$ & 	3/2 & 	0.0	 &   0.00 & 0.00 \\

                              \hline
  \multirow{15}{*}{C}    & 	$^3P_0$	 & 0     &   -3.7 &   0.00 & 	  -1.28 \\
                         &	$^3P_1$  & 0	 &   -4.5 &   0.00 & 	-186.31 \\
                         &	$^3P_1$	 & 1	 &   -1.3 &   0.00 &       -1.93 \\
                         &	$^3P_2$	 & 0	 &   -4.4 &   0.00 & 	 -10.39 \\
                         &	$^3P_2$	 & 1	 &   -1.4 &   0.00 & 	  -2.08 \\
                         &	$^3P_2$	 & 2	 &   0.0 &   0.00 & 	 -28.80 \\
                         &	$^3D_1$	 & 0	 & -131.7 &   0.00 & 	  -7.77 \\
                         &	$^3D_1$	 & 1	 &  -13.6 &   0.00 & 	 -20.77 \\
                         &	$^3D_2$	 & 0	 & -168.0 &   0.00 & 	  -9.79 \\
                         &	$^3D_2$	 & 1	 & -126.7 &   0.00 & 	 -45.86 \\
                         &	$^3D_2$	 & 2     &   0.0 &   0.00 & 	   0.00 \\
                         &	$^3D_3$	 & 0	 & -107.8 &   0.00 & 	  -6.45 \\
                         &	$^3D_3$	 & 1	 &  -28.5 &   0.00 & 	 -36.01 \\
                         &	$^3D_3$	 & 2	 &   0.0 &   0.00 & 	   0.00 \\
                         &	$^3D_3$	 & 3	 &   0.0 &   0.00 & 	   0.00 \\

                              \hline
                              \hline
\end{tabular}
\end{table}

\section{Conclusions}
\label{SEC:CONCLUSIONS}

We have developed and applied constraints on the 1RDM and 2RDM that give us access to multiple total angular momentum ($J$) and  projection of total angular momentum ($M_J$) states in v2RDM-based calculations on atomic systems. Serious issues consistently arise in which $J$ and $M_J$ states that should be energetically degenerate are not, and this problem persists even in a case where 2RDM is exactly ensemble-state $N$-representable. The large energy error we observe for this case results because the 2RDM derives from an ensemble of pure-state 2RDMs that are not themselves associated with good orbital angular momentum quantum numbers. This example clearly demonstrates that pure-state $N$-representability conditions can play a significant energetic role in v2RDM-based optimizations. 

The results presented herein have important consequences for the role of v2RDM theory in quantum chemistry and related fields, such as quantum information science. First, because v2RDM theory fails to recover qualitatively correct results for angular-momentum coupled states in the non-relativistic limit, relativistic extension of the theory will likely be unreliable, at least without the imposition of additional constraints (\textit{i.e.}, pure-state $N$-representability constraints). Second, recent work\cite{McClean18_053020} has demonstrated that the application of ensemble-state $N$-representability conditions can improve the fidelity of quantum simulations on quantum hardware. Our results highlight the danger of reliance on ensemble-state conditions in such applications, as we have demonstrated that an exactly ensemble $N$-representable RDM can exhibit absolute energy errors in excess of 0.1 E$_{\rm h}$. On the other hand, large pure-state $N$-representability errors do not, in general, imply large energy errors. Hence, the computational effort required to restore pure-state $N$-representability to ensemble-state RDMs may not always be justified.

\section{Appendix}
\label{SEC:APPEBDIX}

\appendix
\section{Generating pure-state $N$-representability conditions for high-order RDMs}
In this appendix we review the procedure for constructing pure-state conditions on $p$-body RDMs (where $p\geq 2$) from known 1RDM conditions. The set of pure-state $N$-representable 1RDMs for $N$ electrons in $r$ spin orbitals is denoted $P^{1}_{N, r}$ and is generically defined as 
\begin{align}
P^{1}_{N, r} := \{ \, {}^{1}D \vert \langle {}^{1}\hat{O}_{m} \vert \;{}^{1}D\rangle \geq 0 \;\; \forall \;\; {}^{1}\hat{O}_{m} \in \; B^{1}_{N,r} \, \}
\end{align}
where $B^{1}_{N, r}$ is the set of one-body operators that expose the boundary of $P^{1}_{N, r}$, and $\langle . | . \rangle$ is the operator inner product defining a vector space of size required to specify ${}^{1}D$ and $B^{1}_{N,r}$.  The inner product between a 1RDM and an exposing operator is an equation for a line in the vector space needed to specify $B^{1}_{N,r}$ and ${}^{1}D$. At all points along the the boundary of the pure-state representable set of 1RDMs, the inner product equation describes a bounding hyperplane of the set.  The exposing operators are not uniquely defined with respect to any particular basis and are functions of the eigenvalues of the 1RDM.  Klyachko~\cite{Klyachko06_72} derived these functions for $N$ electrons distributed among $r$ spin orbitals such that the 1RDM is a pure $N$-representable 1RDM.  It is important to note that these conditions are necessary but not sufficient.  

A strategy for determining the set of conditions that should be satisfied by a pure-state $N$-representable 2RDM is to apply the GPCs to a 1RDM that can be obtained from the Gram matrix of a state with one electron removed (added) from (to) an arbitrary orbital.  Consider an arbitrary state $\vert \Psi^{(N \pm 1)} \rangle$ with $N \pm 1$ electrons; we have
\begin{align}
\langle \Psi^{(N \pm 1)} \vert {}^{1}\hat{O}_{m} \vert \Psi^{(N \pm 1)} \rangle \geq \, 0 \;\; \forall \;{}^{1}\hat{O}_{m} \in B^{1}_{N \pm 1, r}
\end{align}
which is a hyperplane constraint with operators exposing the set of pure-state representable ($N \pm 1$)-electron 1RDMs.  Generically, we can define this ($N \pm 1$)-electron state via the removal/addition of an electron from an $N$-electron wave function as
\begin{align}
\label{EQN:NM1_STATE}
\vert \Psi^{(N-1)} \rangle =& \left( \sum_{s} c_{s} a_{s}\right) \vert \Psi \rangle \big / \left(\sum_{p, s}c_{p}^{*}c_{s}\langle \Psi \vert a_{p}^{\dagger}a_{s}\vert \Psi \rangle\right)
\end{align}
and
\begin{align} 
\label{EQN:NP1_STATE}
\vert \Psi^{(N+1)} \rangle =& \left( \sum_{s} c_{s} a_{s}^{\dagger}\right) \vert \Psi \rangle \big / \left(\sum_{p, s}c_{p}^{*}c_{s}\langle \Psi \vert a_{p}a_{s}^{\dagger}\vert \Psi \rangle\right)
\end{align}
We interpret $\vert \Psi^{(N \pm 1)} \rangle$ as the states obtained by adding/removing a particle to/from an arbitrarily defined orbital represented in any basis.  Put another way, we have
\begin{align}
\vert \Psi^{(N-1)} \rangle =& \,\tilde{a}_{u} \vert \Psi \rangle / \langle \Psi \vert \tilde{a}_{u}^{\dagger}\tilde{a}_{u} \vert \Psi \rangle \\
=& \,e^{K}a_{u}e^{-K}\vert \Psi\rangle /  \langle \Psi \vert  e^{K}a_{u}^{\dagger} e^{-K} e^{K}a_{u} e^{-K} \vert \Psi \rangle \\
=& \,\left(\sum_{s}c_{s} a_{s}\right)\vert \Psi \rangle \big / \left(\sum_{p, s}c_{p}^{*}c_{s}\langle \Psi \vert a_{p}^{\dagger}a_{s}\vert \Psi \rangle\right)
\end{align}
and
\begin{align} 
\vert \Psi^{(N+1)} \rangle =& \,\tilde{a}_{u}^{\dagger} \vert \Psi \rangle / \langle \Psi \vert \tilde{a}_{u}\tilde{a}_{u}^{\dagger} \vert \Psi \rangle \nonumber\\
=& \,e^{K}a_{u}^{\dagger}e^{-K}\vert \Psi\rangle /  \langle \Psi \vert  e^{K}a_{u} e^{-K} e^{K}a_{u}^{\dagger} e^{-K} \vert \Psi \rangle \nonumber\\
=& \,\left(\sum_{s}c_{s}^{*} a_{s}^{\dagger}\right)\vert \Psi \rangle \big / \left(\sum_{p, s}c_{p}c_{s}^{*}\langle \Psi \vert a_{p}a_{s}^{\dagger}\vert \Psi \rangle\right)
\end{align}
Here, we have expressed the single particle rotation as
\begin{align}
e^{K} = \exp\left[ \sum_{pq}\kappa_{pq}a_{p}^{\dagger}a_{q}\right] 
\end{align}
with
\begin{align}
    \kappa_{pq} = -\kappa_{qp}
\end{align}
and
\begin{align}
    e^{K}a_{u}e^{-K} = \sum_{s} \left[e^{\kappa}\right]_{u, s}a_{s} =  \sum_{s} c_{s} a_{s}
\end{align}
Another fact we will need is that, given a basis $e^{K}$ or even just a set $\{c\}$, we can define a relationship between of the 1RDM in the ($N \pm 1$)-electron space to the $N$-electron 2RDM in the $N$-electron space as
\begin{align}\label{eq:n1opdm_tpdm}
\langle \Psi^{(N-1)} \vert a_{q}^{\dagger}a_{t} \vert \Psi^{(N-1)} \rangle = \mathcal{N}^{(N-1)}(c)^{2} \sum_{p,s}c_{p}^{*}c_{s}\langle \Psi \vert a_{p}^{\dagger}a_{q}^{\dagger}a_{t}a_{s} \vert \Psi \rangle 
\end{align}
\begin{align}
\label{EQN:1RDM_NP1}
\langle \Psi^{(N+1)} \vert a_{q}^{\dagger}a_{t} \vert \Psi^{(N+1)} \rangle = \mathcal{N}^{(N+1)}(c)^{2} \sum_{p,s}c_{p}c_{s}^{*}\langle \Psi \vert a_{p}a_{q}^{\dagger}a_{t}a_{s}^{\dagger} \vert \Psi \rangle
\end{align}
where we define
\begin{align}
\mathcal{N}^{(N-1)}(c) = \left(\sum_{ij}c_{i}^{*}c_{j}\langle \Psi \vert a_{i}^{\dagger}a_{j}\vert \Psi \rangle\right)^{-1/2}
\end{align}
and
\begin{align}
\mathcal{N}^{(N+1)}(c) = \left(\sum_{ij}c_{i}c_{j}^{*}\langle \Psi \vert a_{i}a_{j}^{\dagger}\vert \Psi \rangle\right)^{-1/2}
\end{align}
as normalization constants.  We now have access to the 1RDM in the $(N \pm 1)$-particle space that allows us to define the natural orbital basis to represent $B^{1}_{N \pm 1, r}$ exposing operators. In Eq.~\ref{eq:n1opdm_tpdm}, the 1RDM for the ($N-1$)-electron state maps directly onto the 2RDM for the $N$-electron state. In Eq.~\ref{EQN:1RDM_NP1}, the 1RDM for the ($N+1$)-electron state maps directly onto the particle-hole RDM for the $N$-electron state, which, of course, can be expressed in terms of the 1RDM and 2RDM for the $N$-electron state.

In order to determine the degree to which the GPC conditions are satisfied, we numerically optimized the coefficients $\{c\}$ of the arbitrary-basis orbital from which the electron is removed (added), as defined by Eq.~\ref{eq:n1opdm_tpdm} (Eq.~\ref{EQN:1RDM_NP1}).  This non-linear optimization is accomplished by maximizing violation of the GPC conditions with respect to the set of coefficients $\{c\}$.  Through $\{c\}$, the 1RDMs for the ($N \pm 1$)-electron states are used to evaluate the GPC conditions.  The sum of violated GPC conditions are then returned as the function value.  We used the COBYLA~\cite{powell1998direct} method implemented in SciPy~\cite{2020SciPy-NMeth} to maximize GPC violation given a 2RDM for an $N$-electron system. Note that this procedure assumes that the coefficients $\{c\}$ are real. Each 2RDM GPC violation reported in this work corresponds to the largest violation observed from a large number of random $\{c\}$ starting points (\textit{i.e.}, $\ge$ 100 independent runs). We have made the code required for this analysis available online.\cite{Rubin22_Zenodo} GPCs for the 
$\wedge^3\mathcal{H}_{8}$,
$\wedge^4\mathcal{H}_{8}$,
$\wedge^3\mathcal{H}_{10}$, $\wedge^4\mathcal{H}_{10}$, and $\wedge^5\mathcal{H}_{10}$ classes have been implemented; these conditions were taken from an online resource\cite{MurataMissingGPCs} (referenced in the preprint version\cite{Klyachko08_arxiv} of Ref.~\citenum{Klyachko08_287}), which no longer appears to exist. Reference \citenum{Klyachko08_287} claims that these conditions are tabulated in the Supporting Information for that work, but this, unfortunately, does not seem to be the case.

{\color{black}It is now worthwhile to compare the derivation above to that presented} in Reference~\citenum{Mazziotti16_032516}, which follows the same principles.  Namely, each approach lifts an $(N-1)$-electron wave function in an arbitrary basis to the $N$-particle space and relates the 1RDM of the $(N-1)$-electron space to the $2$-RDM of the $N$-electron space.  The difference lies in the interpretation of the orbital removal operator.  Reference~\citenum{Mazziotti16_032516} asserts that the 1RDM for the $(N-1)$-electron state comes from a $\wedge^{N-1}\mathcal{H}_{r-1}$ space, but we disagree that the annihilator removes an orbital.  Rather, the ladder operators move between particle-number manifolds of Fock space, given a fixed basis.  Consequently, the violations in the  1RDM for the $(N \pm 1)$-electron state should be considered against GPCs exposing the boundary of $P^{1}_{N \pm 1, r}$ and not $P^{1}_{N \pm 1,r \pm 1}$. 

{\color{black}We conclude by noting a connection between the numerical procedure outlined above and the extended Koopman's theorem (EKT).\cite{Orville75_113,Levy75_549} In EKT, ionized states are parametrized according to Eq.~\ref{EQN:NM1_STATE}, and the expansion coefficients that define these states are chosen such they satisfy the EKT equations
\begin{equation}
    \label{EQN:EKT}
    \sum_{q} \langle \Psi |\hat{a}_p^\dagger [ \hat{H}, \hat{a}_q ] | \Psi \rangle c^k_q = I^{\rm EKT}_k \sum_q \langle \Psi | \hat{a}^\dagger_p \hat{a}_q | \Psi \rangle  c^k_q
\end{equation}
Here, we have introduced an index, $k$, which indicates the existence of multiple solutions to this equation ({\em i.e.}, multiple ionized states), and $I^{\rm EKT}_k$ is the $k^{th}$ ionization potential.
Interestingly, the EKT provides a potentially exact description of the lowest-energy ionized state ($k=0$).\cite{Ernzerhof09_793,Bultinck09_194104} Assuming one has knowledge of the exact ionization potential, any deviation from exactness of $I^{\rm EKT}_0$ could be attributed to representability issues with the 2RDM that determines the left-hand side of Eq.~\ref{EQN:EKT}. Hence, by optimizing $\{c_p\}$ such that they minimize the EKT ionization potential, one gains access to a measure of $N$-representability error that complements that obtained when choosing $\{c_p\}$ to maximize the GPC error. This analysis reaffirms the important role different particle-number manifolds of Fock space can play in assessing the quality and $N$-representablility of RDMs.}

\vspace{0.5cm}

{\bf Supporting Information} Energies for coupled and uncoupled ${}^3$P angular momentum states of a carbon atom obtained from v2RDM calculations.

\vspace{0.5cm}

{\bf Acknowledgments} 
This work was supported by the Army Research Office Small Business Technology Transfer (STTR) program under Grant No. W911NF-19-C0048.

\bibliography{Journal_Short_Name.bib,main.bib}

\end{document}